\documentclass{article}

\usepackage{arxiv}

\usepackage[utf8]{inputenc} % allow utf-8 input
\usepackage[T1]{fontenc}    % use 8-bit T1 fonts
\usepackage{hyperref}       % hyperlinks
\usepackage{url}            % simple URL typesetting
\usepackage{booktabs}       % professional-quality tables
\usepackage{amsfonts}       % blackboard math symbols
\usepackage{nicefrac}       % compact symbols for 1/2, etc.
\usepackage{microtype}      % microtypography
\usepackage{lipsum}
\usepackage{graphicx}
\usepackage{gensymb}
\usepackage{tikz}
\usepackage{tikzscale}
\usepackage{amsmath}
\usetikzlibrary{arrows.meta}
\usetikzlibrary{shapes.geometric}
\usetikzlibrary{positioning}
\usetikzlibrary[arrows,decorations.pathmorphing, decorations.markings,backgrounds,positioning,fit,petri,calc]
\usetikzlibrary{calc}
\usepackage{subfig}

\usepackage{algorithmic}
\usepackage{algorithm}

\graphicspath{ {./images/} }

\title{Deep Unsupervised Identification of Selected SNPs between Adapted Populations on Pool-seq Data}

\author{
 Julia Siekiera \\
  Department of Computer Science\\
  Johannes Gutenberg University\\
  Mainz, 55122 \\
  Germany\\
  \texttt{siekiera@uni-mainz.de} \\
  \And
 Stefan Kramer \\
  Department of Computer Science\\
  Johannes Gutenberg University\\
  Mainz, 55122 \\
  Germany\\
  \texttt{kramer@informatik.uni-mainz.de}
}

\begin{document}
\maketitle
\begin{abstract}
The exploration of selected single nucleotide polymorphisms (SNPs) to identify genetic diversity between different sequencing population pools (Pool-seq) is a fundamental task in genetic research. As underlying sequence reads and their alignment are error-prone and univariate statistical solutions only take individual positions of the genome into account, the identification of selected SNPs remains a challenging process.
Deep learning models like convolutional neural networks (CNNs) are able to consider large input areas in their decisions. We suggest an unsupervised pipeline to be independent of a rarely known ground truth.
We train a supervised discriminator CNN to distinguish alignments from different populations and utilize the model for unsupervised SNP calling by applying explainable artificial intelligence methods. Our proposed multivariate method is based on two main assumptions: 
We assume (i) that instances having a high predictive certainty of being distinguishable are likely to contain genetic variants, and (ii) that selected SNPs are located at regions with input features having the highest influence on the model's decision process.
We directly compare our method with statistical results on two different Pool-seq datasets and show that our solution is able to extend statistical results. 
\end{abstract}

\section{Introduction}
The investigation of populations under different environmental pressures provides insights into evolutionary processes.
A cost- and time-effective approach to generate population genomic data is sequencing population pools (Pool-seq) \cite{Futschik_Schloetterer}. By pooling the DNA of different individuals in a population, a sequence alignment can subsequently be generated with the aid of a reference genome. The corresponding allele frequencies can be estimated from the alignment.
To further compare natural or experimental populations, genomic studies investigate especially genetic variants like single nucleotide polymorphisms (SNPs) of the generated sequence alignment.

%Related work:
In general, the identification of selected SNPs is sequentially done in two steps. First, all variable positions of the genome are identified during SNP calling taking sequencing errors under consideration. Second, the SNPs are are further analyzed by empirical or model based approaches to find selected SNPs.

Usually simple univariate statistics considering only the current position of the sequence alignments are applied during both steps.
Variant calling for Pool-seq data is addressed by SNVer \cite{SNVer}, that interprets variant calling as hypothesis testing problem or CRISP \cite{crisp}, that discovers SNPs by the use of contingency tables and the consideration of sequencing errors. The R-library PoolSeq \cite{poolSeq} is an example of a model based approach to find selected SNPs by the estimation of selection parameters. But PoolSeq can only be applied on replicated time series data of Evolve and Resequence studies. A commonly used software tool to compare allele frequencies between different populations is Popoolation2 \cite{popoolation2}. The tool provides genome-wide analysis on different window sizes supported by statistics like the allele frequency differences, the fixation index or  Fisher's exact test to find selected SNPs empirically. However, the sequential examination of individual positions in the sequence alignments omits additional knowledge from adjacent regions.

Convolutional Neural Networks (CNNs), on the other hand, are very successful in pattern recognition of images \cite{image_reco} or sequence data \cite{speech_reco} to make inferences.
By considering each feature in the context of its neighbours, they take local properties of the data into account. In more recent research, CNNs are also used in population genetic inference \cite{popul_inference} and are successfully applied to variant and indel calling in aligned next-generation sequencing read data \cite{deepVariant}. 
The supervised CNN DeepVariant \cite{deepVariant} interprets the reference genome and aligned reads with their quality scores as image to call variants. However, the architecture is limited to variant calling on single individuals.
For the comparison of different populations from Pool-seq data, no deep learning solution exists to find selected SNPs. 
To obtain a multivariate model for selected SNP identification on Pool-seq data, we aim to develop a CNN architecture that is independent from any prior knowledge, as well-known selected SNPs are rare for many individuals or based only on univariate statistical solutions.
We propose to solve this unsupervised task indirectly with the aid of a supervised discriminator model and the tracing of the most significant input features.
To find significant SNP positions between two populations we make the following assumptions: 
\begin{itemize}
    \item A genomic region of two populations is likely to contain SNPs if our auxiliary model is able to discriminate them.
    \item The most influential input features in the decision process indicate the SNP positions.
\end{itemize}
The remainder of the paper is organized as follows: In section \ref{methods}, we describe the methods used in our suggested selected SNP identification pipeline. In section \ref{evaluation}, we explain our experimental setup with different datasets and compare our solution with the SNPs found by univariate statistical approaches. The final conclusion is given in section \ref{conclusion}.
\section{Methods}
\label{methods}
\subsection{Data preprocessing}
\label{data_preprocessing}
% Pfenninger reprocessing:
We work with already investigated Pool-seq data that includes the environmental adaptation of different populations provided by Pfenninger et al. \cite{pfenninger_poecilia} and Pfenninger and Foucault \cite{pfenninger_foucault}. To integrate this data in our pipeline, we adopt the preprocessing from raw reads to the final alignment.
This preprocessing includes, amongst others, controlling the raw reads sequencing quality and trimming. Reads were aligned to the reference genome and indel regions were removed from the alignment. The exact procedure and the programs used are described in Pfenninger et al. \cite{pfenninger_poecilia} and Pfenninger and Foucault \cite{pfenninger_foucault}. The resulting alignments contain the base counts for every position in the gene.

% My prepocessing:
To generate the classifier's input instances, we cut the alignment into separate regions based on the positions of protein-coding genes. In order to find a good trade-off between utilizing the entire information content of all protein-coding genes and the input size limitations of the classifier, we set the maximum gene length to the length that is required to be able to utilize 80\% of the genes without information loss. Note that we also take the segments between protein-coding genes as input under consideration. We define the length of the alignment instances $region\_length$ as the maximum gene length.
With the dimension $region\_length * 4$ (where 4 is the number of bases of the alignment) the resulting instances are large enough to generate an expressive model, but small enough to limit the training time. Missing or removed entries are padded with 0 in the instances. As we compare two corresponding instances of different populations in our model, we delete all counts of instance $x_{i,j}$ at position $j$ that have a local coverage of zero in $\widetilde{x}_{i,j}$ and vice versa, to get a more comparable pairwise input. Finally, we divide each count by the local coverage to consider only the allele frequency as input entries (often referred to as features). Although we increase the information loss with this procedure, this step is necessary to prevent the network from trivial solutions based on the alignments' coverage.

\subsection{Alignment discrimination using CNNs}
CNNs are pattern recognition algorithms which are able to interpret high-dimensional input data $X$ by the optimization of an objective function. In the case of supervised machine learning, models like CNNs learn from a set of labeled instances $(x,y)\in X\times Y$ and should be able to predict the labels $y\in Y$ of unseen instances. In contrast, unsupervised models aim to find patterns in the dataset that deviate from unstructured noise without the use of pre-existing target values $y\in Y$. For the interested reader, we recommend to study LeCun et al. \cite{lecun2015deeplearning} for a more detailed overview of the functionality of supervised deep neural networks (DNNs) like CNNs.

To find significant SNPs between two populations by an unsupervised approach, we designed a supervised discriminative CNN in a first step. We expect that a discriminative model, which is able to distinguish pairwise alignments of the same genomic region from different populations, could also be used to find SNPs in a second unsupervised step. Accordingly we assume the location of selected SNPs at positions of features that have an increased influence on the model's decision process. To generate labeled data, we arrange the pairwise input as follows. Let $x_{i,w_{n}}$ be the $i$'th instance of population $w_{n}$ and $x_{i,a_{n}}$ the $i$'th instance of population $a_{n}$. The model always compares populations $(w_{n},a_{n})$ whose SNPs we are interested in, with $(w_{n},a_{n})\in V=\{(w_{0},a_{0}),(w_{1},a_{1}),..\}$, where $w_{n}$ represents the index of a wild-type population and $a_{n}$ the index of an adapted population. The input $(x_{i,w_{n}},x_{i,a_{n}})$ and $(x_{i,a_{n}},x_{i,w_{n}})$ are then labeled as 0 and 1, respectively. This task is equivalent to determining the order of the pairwise input alignment instances.

\begin{figure*}[t!]
\centering
\resizebox{0.7\textwidth}{!}{
 \includegraphics[width=1.0\textwidth]{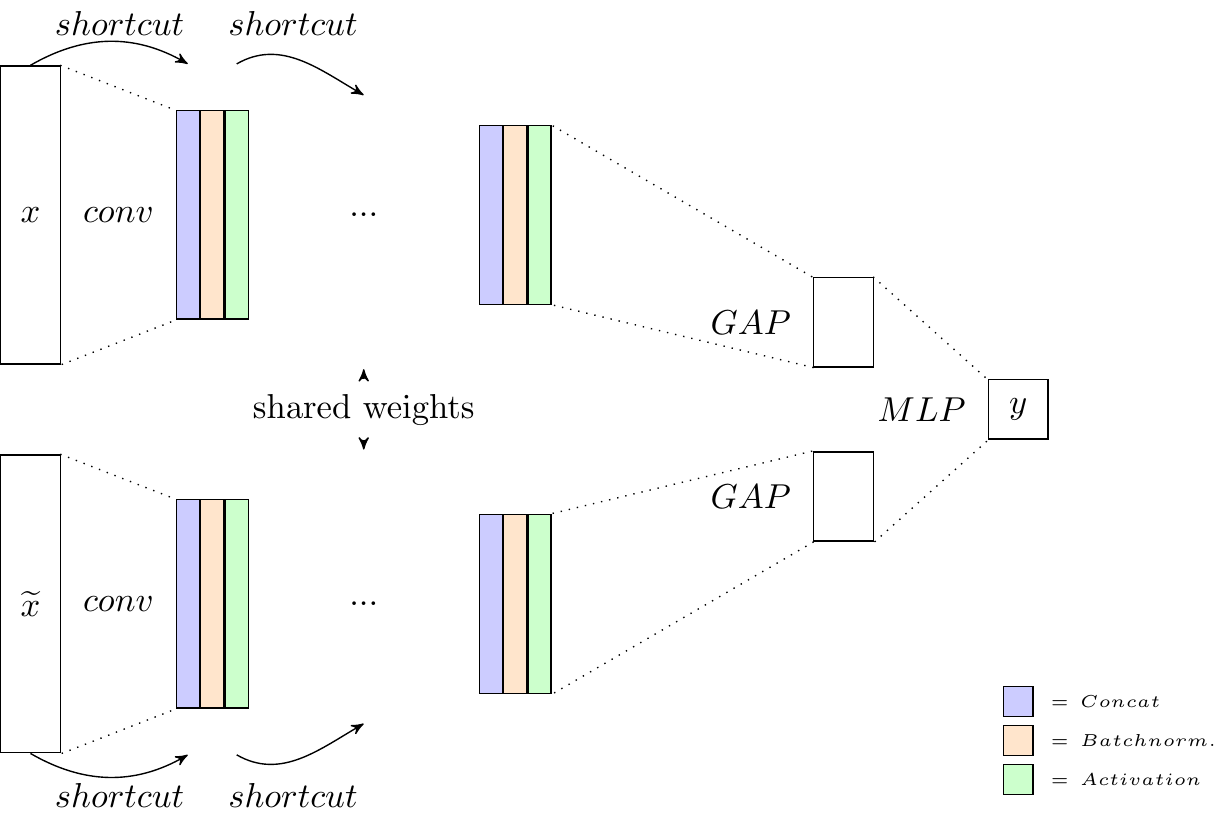}
}
\caption{Alignment discriminator: The two input alignments are aggregated with several convolutional (conv) layers and Resnet shortcuts. The concatenated GAP of the CNN is applied to a MLP with softmax activation to obtain the final output.}
\label{fig:CNN}
\end{figure*}

%To summarize, the model gets two alignments of the same genetic region and has to predict their order.
% TODO explain in more detail!
In general, our architecture consists of two parts. The first one compresses the input representation of each instance from the pairwise input by the application of several convolutional (conv) layers and the second one distinguishes the two compressed representations with a single multi-layer perceptron (MLP). An overview of the architecture is given in Figure \ref{fig:CNN}.
Our model reduces each of the two input instances by the use of several 1d convolutions with max-pooling Residual Network (ResNet) shortcuts \cite{ResNet}. ResNet shortcuts are especially beneficial in networks with increased depth, as they give the network the opportunity to skip parts of the stacked network architecture. Although our architecture contains a limited number of conv layers, they turned out to be effective during our experiments, as they lead to a faster training with increased accuracy.
To use a higher learning rate, which also reduces the training time, we include batch normalization \cite{batchnorm} after every conv layer and subsequently apply the activation. Batch normalization standardize mean and variance of each network unit in order to stabilize the learning process.
The two output instances of the CNN part are aggregated over the dimension of the length with global average pooling (GAP) and concatenated. This layer is applied to the MLP with softmax activation to receive the output $y$. The network uses stochastic gradient decent as optimizer and is trained with cross-entropy loss as objective function.

\subsection{Identification of selected SNPs}
We propose a pipeline with two selection steps, to extract the most crucial SNPs between two alignments. First, based on the model's predictive discrimination certainty we extract instances that are likely contain selected SNPs. Second, we specify the exact SNP positions of each candidate based on the influence of the different input features on the model's discriminative decision.

During the first selection step we assume that if the model is able to distinguish two input regions, those regions should differ at some positions significantly. 
Therefore, we use the model's discriminative predictions as certainty measure for each region:
\begin{equation}
\label{eq:certainty}
certainty=\max\limits_{y\in Y} p(y|x)  
\end{equation}
Instead of just taking the probabilities of the true class labels under consideration, we use the maximum of the two possible outcomes. This approach is advantageous, since an increased probability of the reversed (false) order also indicates a differentiation in the alignments. Therefore, we end up with an equivalent measure that is successfully applied in active uncertainty learning \cite{uncertainty_al}.

We sort the instances according to their certainty and assume the first $\lambda_{1}$\% with the highest certainty containing significant variants. The selected instances are further investigated in the second step. 
In this step we calculate the influence of each input feature on the model's prediction.
In image recognition, Gradient-weighted Class Activation Mapping (Grad-CAM) \cite{Grad_CAM} is used to obtain visual explanations of CNN predictions. We believe that this method could also be applied to our sequence data to measure feature importance. The approach combines the following two methods:
\begin{itemize}
    \item First, high-resolution, but not class-discriminative guided backpropagation \cite{Backpropagation}. Guided backpropagation calculates the gradient of the output with respect to the input $\frac{\partial f(x)}{\partial x}$ and sets all negative gradients to zero during backpropagation. 
A large gradient implies that the output neuron would not be activated, if the input feature at this position would be changed by some small $\epsilon$. Therefore this approach can be used as indicator of input feature influence.
\item Second, the class-discriminative Class Activation Maps (CAM) that combine the summarized feature importance from the GAP layer and the weights of the MLP. The CAM of a certain class $c$ is calculated by: $M_{c}(x,y)=\sum\limits_{k}w_{k}^{c}f_{k}(x,y)$, where $w_{k}^{c}$ represents the weight from the $j$'th GAP feature to the output neuron of class $c$ of the fully connected layer. And $f_{k}(x,y)$ represents the $k$'th GAP feature. As only the values of the highly aggregated GAP are used, CAMs do not have sufficient resolution to directly map the measure to the input. Instead, the results need to be resized to the input dimension, using a method like bilinear resizing.
\end{itemize}

For the same reason as in step 1, we are not only interested in the information of the true class, but also in the information of the false class. We therefore utilize the sum over the CAM results. As we constructed a binary classification task, we can additionally assume that high negative CAM values indicate an increased influence of the opposite class. Hence, we use the absolute CAM values.
Like Grad-CAM, we combine the results of both methods by point-wise multiplication, to generate our final input feature importance as heatmap:
\begin{equation}
    \label{eq:heatmap}
    heatmap=guided\_backprop(x,y)*resize(\sum\limits_{c\in Y}|M_{c}(x,y)|)
\end{equation}{}
To provide a sufficient allele frequency at the predicted SNPs, we ensure that the unified base coverage of the two alignments differs by $\lambda_{2}$\% at all potential SNPs positions. These potential positions are sorted according to the resulting heatmap and the positions with the highest values are selected as SNP. The complete process is summarized in algorithm \ref{algo_summary}.

\begin{algorithm}[t]
  \caption{$\mbox{SNP selection pipeline}$}
  \begin{algorithmic}[1]
  \REQUIRE{Alignment instances $X$}
  \ENSURE{Trained Classifier $C$, potential SNP positions $P$, SNP heatmap rankings $H$, selected SNP positions $S$}
    \STATE Train classifier: $C\leftarrow Train(X)$ 
    \STATE Compute certainty score (Equation \ref{eq:certainty}) for all $x\in X$ 
    \STATE Select $\lambda_{1} \%$ of the most certain instances: $E\subset X$
    \FOR{$x_{i}\in E$}
        \STATE Compute potential SNP positions $p$ with $\lambda_{2}\%$ base differentiation
        \STATE $P\leftarrow P\cup p$
        \FOR{$x_{i,j}\in P$}
            \STATE Compute heatmap ranking $h_{i,j}$ (Equation \ref{eq:heatmap})
            \STATE $H\leftarrow H\cup h_{i,j} $
        \ENDFOR
    \ENDFOR
    \STATE Select $\lambda_{3}\%$ of $P$ with highest ranking in $H$: $S\subset P$
  \end{algorithmic}
      \label{algo_summary}
\end{algorithm}
\begin{table}[h!]
 \caption{Dataset parameters}
  \centering
  \begin{tabular}{lll}
    \toprule
    Parameter     & Dataset 1     & Dataset 2 \\
    \midrule
    $region\_length$ & 20000 bp  & 8000 bp    \\
    Training set size     & 44100 & 17900     \\
    Validation set size     & 4900       & 2000  \\
    \bottomrule
  \end{tabular}
  \label{tab:table_configs}
\end{table}
\begin{figure*}[t!]
\centering

\subfloat{
\resizebox{0.3\textwidth}{!}{
\includegraphics[width=1.0\textwidth]{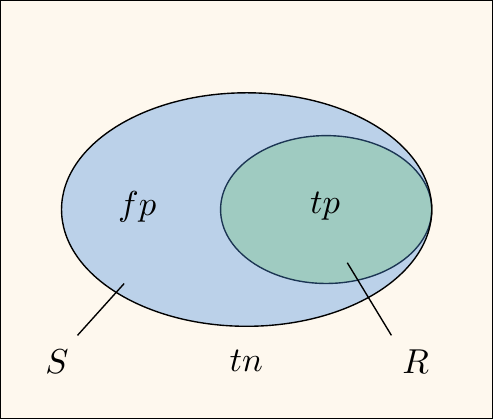}

}}
\subfloat{
\resizebox{0.3\textwidth}{!}{
\includegraphics[width=1.0\textwidth]{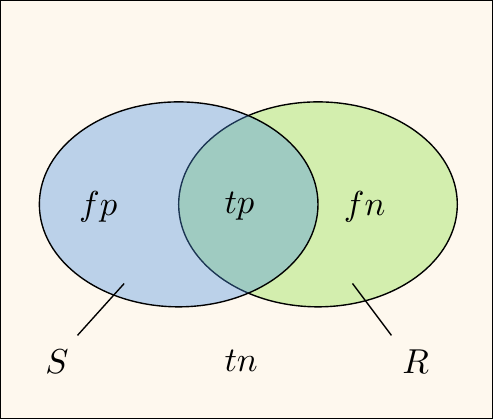}

}}
\caption{Selected elements $S$ and relevant elements $R$ in binary classification with corresponding true positives $tp$, false positives $fp$, true negatives $tn$ and false negatives $fn$: An optimal selection (left) is assumed to contain all relevant (univariate) results and additional (multivariate) results which can not be found with univariate methods. A less optimal selection (right) contains many false negatives $fn$.}
\label{fig:test}
\end{figure*}
\section{Evaluation}
The decision whether the additional multivariate SNPs actually indicate additional phenotypic adaptations in the population depends on the knowledge of the affected gene functionality and their causal relationships with the observed phenotypic traits. However, to gain these relationships, numerous extensive studies would have to be conducted. These studies would require further complex and time consuming investigations of at least one group of adapted populations, which are not feasible within the scope of this research. 
%Good example with time + money/personal requirements would be nice
For this reason, we evaluate the quality our multivariate results with different indirect methods. This includes the comparison with univariate SNPs, the verification of multivariate dependencies and the identification of the most common patterns that are expected to influence the decision process.
\label{evaluation}
\subsection{Experimental setting}
As mentioned in section \ref{data_preprocessing}, we use the already preprocessed datasets provided by Pfenninger et al. \cite{pfenninger_poecilia} and Pfenninger and Foucault \cite{pfenninger_foucault} for the evaluation.
Dataset 1 contains the alignments of four {\em Poecilia mexicana} populations. These populations are habitated in two drainages, with one population per drainage having adapted to hydrogen sulphide containing water. We train the model on this data with all possible combinations of adapted and non adapted populations and detect selected SNPs between the populations habitated in the Puyacatengo drainage.

Dataset 2 includes the alignments of an evolve and resequence study of {\em Chironomus riparius} populations. The data includes alignments of a field population $n_{0}$ and their descendants bred in the laboratory at different temperatures (two populations at 14\degree C, 20\degree C and 26\degree C) for two years. In this case, we train the model to distinguish the field population $n_{0}$ from the others and show SNP calling between $n_{0}$ and the second 26\degree C population.

%Model parameters
During evaluation, we use the following parameter configuration:
4 ResNet layers with stride 2, a filter size of 5 and 32, 64, 128, 256 as number of filters for the individual layers, stochastic gradient decent as optimizer, 0.1 as learning rate, a batch size of 50 and ReLu as activation function. The individual configurations of dataset 1 and 2 are listed in table \ref{tab:table_configs}.

Through our data construction, the labels are discrete uniform distributed. Therefore, accuracy is well-suited to measure the model's discriminative performance.
We stop training, if the accuracy of the validation set is not improving for at least 20 training iterations and take the model with the best validation accuracy.

% "ground truth"
For an ultimate evaluation, ground truth SNP positions would be necessary. However, this leads to the fundamental problem of evaluation, as the existence of a corresponding method would be required. Additionally, we are not able to generate artificial data that is capable to map the hidden effects of genome adaptation, which should be discovered by the CNN in the first place.
Thus, we compare our results with the outlier SNPs found by Pfenninger et al. \cite{pfenninger_poecilia} and Pfenninger and Foucault \cite{pfenninger_foucault}.
These are the results of univariate statistics considering only the base counts of one current position. For comparison, we assume these SNPs as ground truth. 
The ground truth of dataset 1 was chosen empirically as the top 1\% outliers based on the $F_{ST}$-distribution \cite{pfenninger_poecilia}. In contrast to this, the ground truth of dataset 2 was selected based on a genetic model approach \cite{pfenninger_foucault}, possibly leading to results of higher quality.
Of course, the results cannot be compared directly with each other, as they are generally based on different assumptions. However, we assume that a multivariate method should extend the solution of an univariate method, and a comparison should show to which degree our results complement univariate statistics. 
Figure \ref{fig:test} depicts the comparison of an optimal SNP selection, in which the multivariate results form the superset of the univariate results, with a usually less optimal selection.

As the model naturally has an increased discriminative performance on the training set, we assume that the model is able to extract more information on this data for the SNP detection task. Because the training set is only used during our auxiliary task,  the discriminator generation, but not used to find the individual SNP positions, we are able to evaluate our method on it.% During the following evaluation, we first show the effects of the certainty selection and subsequently the results of the entire pipeline.
\subsection{Results on candidate identification using certainty}
\begin{figure*}[t]
    \includegraphics[width=1.0\textwidth]{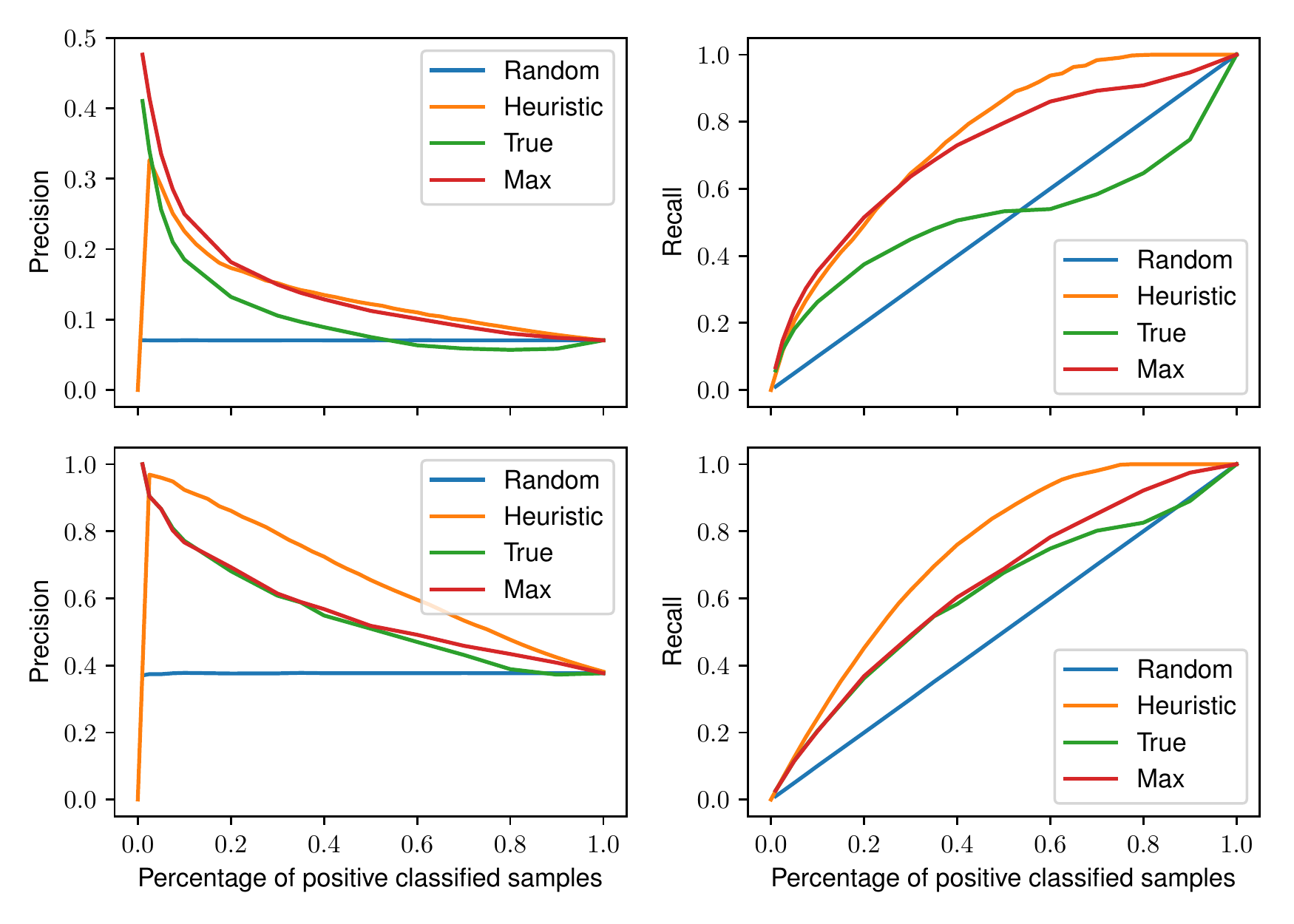}
    \caption{The two upper plots show precision and recall of dataset 1 and the lower ones the results of dataset 2 on the candidate identification task. We plotted the measures against the percentage of positive classified instances to find a suitable $\lambda_{1}$ achieving for both recall and precision high values. $True$ chooses instances based on the true class model prediction and $Max$ selects instances based on equation \ref{eq:certainty}.}
    \label{fig:stage1}
\end{figure*}
During the first evaluation step, we focus on the extraction of instances containing selected SNPs based on the model's predictive certainty. We interpret this task as classification problem. Each instance containing at least one selected univariate SNP is labeled as positive, and negative otherwise. We compare our method against a baseline that picks positive instances randomly. Since we observed in our experiments that the certainty generally tends to choose instances with a higher number of potential selected SNPs, we additionally show the results of a simple heuristic that selects instances based on the number of potential selected SNPs. Figure \ref{fig:stage1} shows the results of recall and precision against the percentage of positive predicted instances $\lambda_{1}$ of both datasets. 
Taking instances with a high predictive certainty generally increases precision on the datasets. 
Since we pursue the goal of supplementing univariate statistical methods, we search for a superset of the ground truth. Thus, we focus on optimizing the recall, as it measures how many relevant solutions are selected. 
The heuristic achieves a similar or better recall than the certainty. Nevertheless, we prefer using the certainty, as this selection fits to the assumptions made in the second evaluation step.

As expected, considering the maximal prediction and, not only the outcome of the true class, has a beneficial effect on both precision and recall. This effect is amplified in dataset 1. A possible explanation could be provided by the discriminative accuracy. In the comparison of the two considered populations, the model reaches an accuracy of 0.36 and 0.79 on dataset 1 and 2, respectively. As the predicted order of the input is switched in dataset 1, the certainty is stronger if considering both model outcomes.
\subsection{Results on selected SNP identification}
\begin{figure*}[t]
    \includegraphics[width=1.0\textwidth]{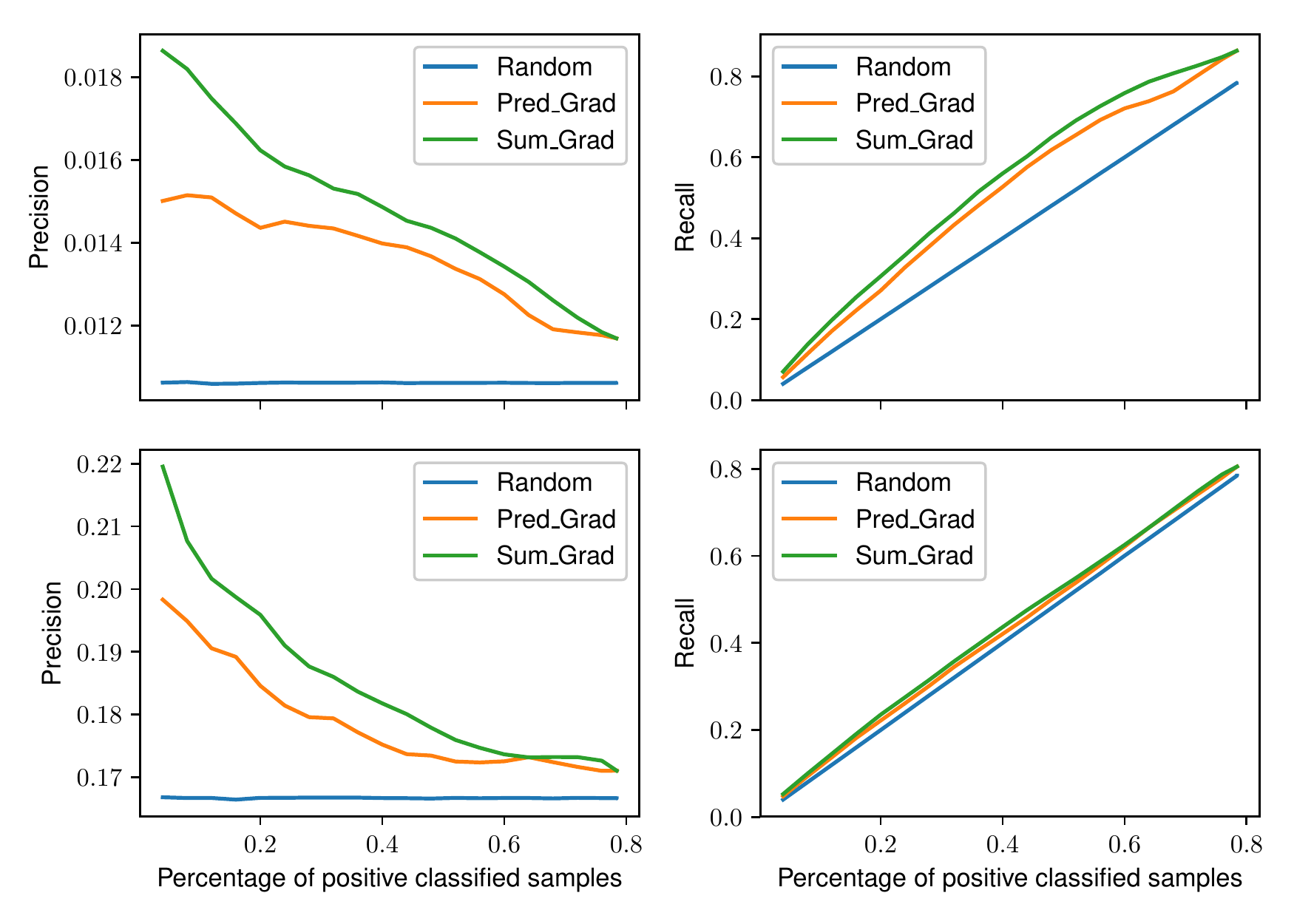}
    \caption{The two upper plots show precision and recall of dataset 1 and the lower ones the results of dataset 2 on the SNP calling task. We plotted the measures against the percentage of positive classified instances $\lambda_{3}$. $Pred\_Grad$ selects SNPs based on the Grad-CAM of the predicted class and $Sum\_Grad$ uses equation \ref{eq:heatmap}. }
    \label{fig:stage2}
\end{figure*}
In the second evaluation step, we choose the SNP positions by the inspection of the most important input features. We show the evaluation on the positive predicted instances at a $\lambda_{1}$ of 50\%. 
Like in the previous section, this task is analyzed as classification problem of the univariate positions. The total set of potential results is represented by the possible SNP positions $P$ based on $\lambda_{2}$ with 15\% base differentiation. By the choice of this low  $\lambda_{2}$, $P$ also covers all possible univariate results. In dataset 1, $|P|=992483$ and $|R|=10539$, the number of univariate positions ($P\subset R$), is relatively small,  since $R$ is chosen as the top 1\% outliers according to $F_{ST}$-distribution. %naive snps 1694716
%According to the different genome length and selection method 
Dataset 2, however, has $|P|=109551$ and $|R|=18256$. %naive snps 597240
%TODO: lookup the numbers!!!
The results are compared to a baseline picking random positions out of all possible SNP positions $P$. % to investigate only the effect of the network performance.
Figure \ref{fig:stage2} shows precision and recall of both datasets. Additionally, we provide in table \ref{tab:absolute_values} of the appendix the absolute values of the detected SNPs and the affected genes for different choices of $\lambda_{3}$. 

Generally, the precision of the random baseline and our pipeline depends on the size of $R$ and the selection of $\lambda_{2}$. As we choose a low $\lambda_{2}$, the precision is accordingly low. Selecting SNPs with an increased heat generally leads to a higher precision, but the effect is not as strong as in evaluation step 1. 
Like in the first step, we observe an improvement of our summarized Grad-CAM version in comparison to the Grad-CAM of the predicted class, and the effect is stronger on dataset 1.
For both datasets, we reach a higher recall than the random baseline, indicating that our unsupervised pipeline is able to extract useful information from our model. 
The success of the discriminating model implies that the alignments contain patterns that the CNN is able to find.

\subsection{Multivariate properties}
\begin{figure*}[t]
    \centering
    \includegraphics[width=0.7\textwidth]{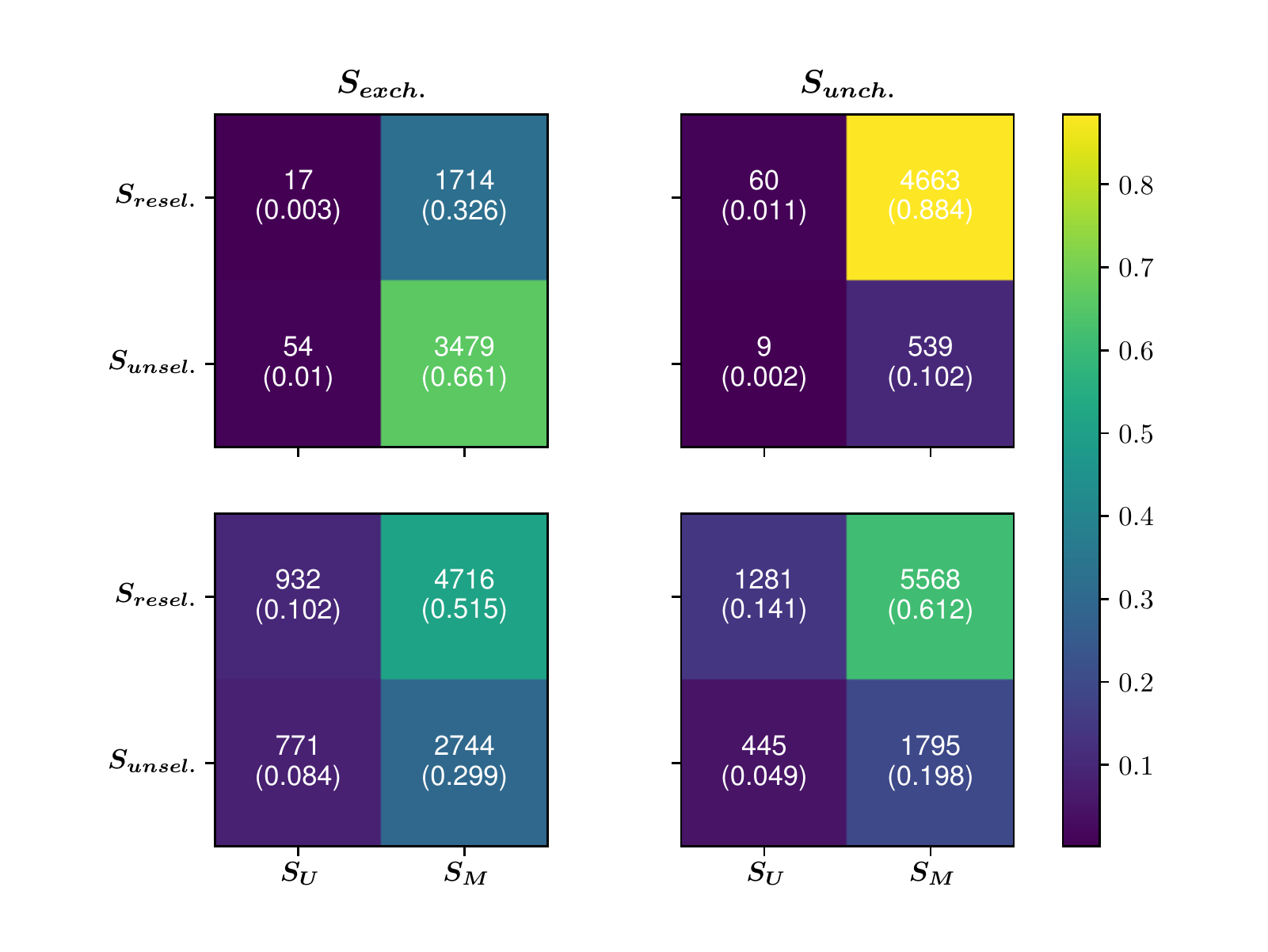}
    \caption{Average results of the randomized SNP exchange with absolute and relative frequencies. The relative values of a plot are the percentages of the 4 different subsets in the exchanged and non-exchanged case. A SNP is denoted as re-selected if the pipeline choose the SNP again after re-execution and as unselected else. A SNP is denoted as univariate if the position was also chosen by the corresponding univariate method and as multivariate else. The two upper plots show the results of dataset 1 and the lower plots show the results of dataset 2.}
    \label{fig:snp_exchange}
\end{figure*}
To demonstrate the existence of patterns influencing the SNP selection pipeline, we investigate the multivariate properties of the SNPs found by our method.
%In the following evaluation step the multivariate properties of the SNPs we found are investigated. 
We assume that multivariate properties are especially present, if after the exchange of a SNP position between the two compared populations, this position is probably not selected by the re-executed algorithm again. This procedure decomposes the multivariate relationships, but does not change the allele frequency. Therefore, the results of univariate methods would remain unchanged on this data, whereas the results of a multivariate method should change. 
For a detailed analysis with both datasets, we randomly exchange the positions of the two populations in 50\% of the initially found SNPs $S$, called $S_{exch.}$, and let the other positions $S_{unch.}$ remain unchanged. After the re-execution of our pipeline, we divide $S$ into the re-selected SNPs $S_{resel.}$ and the unselected SNPs $S_{unsel.}$. 
%The sets are further subdivided into positions that have been exchanged or not exchanged and positions $S_{U}$ that have an intersection with the univariate ground truth.
Furthermore, we distinguish between SNPs $S_{U}$ that have an intersection with the univariate ground truth and SNPs $S_{M}$ that were exclusively found by our pipeline ($S=S_{exch.}\cup S_{unch.}=S_{resel.}\cup S_{unsel.}=S_{U}\cup S_{M}$). 
We repeat this procedure 10 and 5 times for dataset 1 and 2, respectively. In figure \ref{fig:snp_exchange}, we present the average number of positions in the corresponding set intersections as a heatmap of relative frequencies for both datasets. 

Additionally, we examine the results with two different comparison functions. The first function focuses on % 
$S_{M}$. We compute the percentage difference of re-selected and exchanged SNPs and re-selected and non-exchanged SNPs with:
%$\frac{\#re-selected\_exchanged}{\#re-selected\_exchanged+\#unselected\_exchanged}<\frac{\#re-selected\_unexchanged}{\#re-selected\_unexchanged+\#unselected\_unexchanged}$
\begin{equation*}
f_{1}(S)=
\frac{|S_{resel.}\cap S_{unch.}\cap S_{M}|}{|S_{resel.}\cap S_{unch.}\cap S_{M}|+|S_{unsel.}\cap S_{unch.}\cap S_{M}|}- 
\frac{|S_{resel.}\cap S_{exch.}\cap S_{M}|}{|S_{resel.}\cap S_{exch.}\cap S_{M}|+|S_{unsel.}\cap S_{exch.}\cap S_{M}|}
\end{equation*}
to verify that the multivariate method is not able to re-select a part of the modified positions. As expected, we obtain the positive result of 0.566 and 0.124 for dataset 1 and 2, respectively, which confirms our assumption. 
%show the calculated values here: 
%0.6321<0.7561(midge) -39.694149535123245 p-value 2.4066333049013926e-06
%(fish) 0.330131<0.89631 #t-statistic: -393.4312578455024 p-value 2.2539681766164173e-20
Due to the change of the overall ranking, positions that have not been exchanged are also discarded after the pipeline re-execution. However, their percentage is lower than that of the exchanged positions.

The second function:
\begin{equation*}
f_{2}(S)=\frac{|S_{resel.}\cap S_{exch.}\cap S_{U}|}{|S_{resel.}\cap S_{exch.}\cap S_{U}|+|S_{unsel.}\cap S_{exch.}\cap S_{U}|}-\frac{|S_{resel.}\cap S_{exch.}\cap S_{M}|}{|S_{resel.}\cap S_{exch.}\cap S_{M}|+|S_{unsel.}\cap S_{exch.}\cap S_{M}|}.
\end{equation*}
focuses on the exchanged SNPs and investigates the relation between $S_{M}$ 
%that were exclusively found by the multivariate method (called multivariate) and SNPs
and $S_{U}$.
%that were also detected by the univariate method (called univariate). 
The percentage of re-selected $S_{M}$ should be equal or even smaller than the percentage of re-selected $S_{U}$. %For dataset 1 no significant conclusion can be made as the corresponding $S_{U}$ is comparably small and causes a high variance. 
%The subsets of dataset 2 have a lower variance and 
%Both datasets contain the relation:
%$\frac{\#re-selected\_univariat}{\#re-selected\_univariat+\#unselected\_univariat}<\frac{\#re-selected\_multivariat}{\#re-selected\_multivariat+\#unselected\_multivariat}$
Dataset 1 and 2 give the negative result of -0.093 and -0.085, respectively. 
%midge(0.54744<0.6321) #t-statistic: -26.015472507576003 p-value 1.2970554880880723e-05
%fish(0.2374<0.33013) t-statistic: -5.6848788029177335 p-value 0.0003000035379154001
This result could initially seem rather unexpected, since $S_{M}$ should be affected the same or even stronger by the decomposition of multivariate properties than $S_{U}$.
An explanation can be provided by the consideration of allele frequency differences.
Multivariate connections are more likely to be interrupted at positions with higher frequency differences between the populations. Allele frequencies differ less on $S_{M}$ than on $S_{U}$, due to the choice of the low $\lambda_{2}$ that controls frequency differences. The position selection according to Grad-CAM seems to be more influenced by multivariate dependencies than by the frequency differences of the current position. Hence, positions that are also detected by the univariate method are more likely to be not selected by our pipeline after exchange.

We repeated this experiment in a modified version by changing 5 random positions in a 1000 bp area of a selected SNP in both populations.
In this version, the multivariate relationships are interrupted by the random change of a few positions, and the percentage of re-selected SNPs is therefore not dependent on the allele frequency difference of the corresponding positions.
While functions $f_{1}$ still returns a positive result, $f_{2}$ is zero in this setting. Thus,  $S_{U}$ is influenced in the same way as $S_{M}$ by multivariate changes.
\subsection{Recognition of influencing patterns}
While univariate methods detect non-random allele frequency differences during the declaration process of selected SNPs, our multivariate method depends on the CNN's recognition of multiple patterns to distinguish the alignments of two populations. CNNs typically act as black-box whose decisions are not directly traceable. 
To find the patterns that our CNN probably makes use of, we exploit an empirical feature extraction approach with the aid of a random forest classifier. This approach shows the most frequent patterns around regions of selected SNPs that differ from other SNP regions and may be detected by the CNN.
In this experiment, we focus on the 1000 bp regions $P_{region}$ around all possible SNP candidates and encode the alignments as string with the alphabet $\Sigma=\{A, C, G, T, a, c, g, t, O\}$. Undefined positions are encoded with $O$, SNPs with lower case letters and all remaining positions with upper case letters. $A/a, G/g, T/t$ and $C/c$ represent the base with the maximum allele frequency at the corresponding position. 
We compare the pattern frequencies of the regions with the selected SNPs $S_{region}$ found by us and the pattern frequencies of all SNP regions $P_{region}$. %($S_{region} \subset P_{region}$). 
All occurring patterns and their frequencies in each set are determined, with each pattern being counted once per region. The string miner of Dhaliwal et al. \cite{5645631} is applied for the efficient discovery of all patterns with a minimum frequency of 10\% in $S_{region}$ and a maximum frequency of 30\% in $P_{region}$.

We aim to train a model that is able to distinguish the regions based on the pattern frequency. Therefore we separate a maximum of 1000 patterns with the largest frequency difference between the two sets that have a higher frequency in $S_{region}$ and exclude all patterns longer than 15 bases, as they include mostly undefined positions. In the next step we transform each region into a feature vector that describes how many times pattern $i$ is present in a region. The corresponding binary target value is 1 if the encoded region is an element of $S_{region}$ and 0 otherwise. The feature vectors with target values are separated in training and test set to train a random forest classifier that distinguishes regions with and without selected SNPs. %Random forest classifiers provide the advantage of a natural interpretable model structure of the individual trees. We use this advantage to separate the patterns with the largest influence on the classifiers decision process based on the Mean Decrease in Impurity (MDI) \cite{breiman2001random}.\\
Random forest classifiers provide the advantage of fast training. 
With a fast classifier and the reduced number of considered patterns, we are able to calculate the feature importance empirically. 
Feature importance could be directly measured based on the Mean Decrease in Impurity (MDI) \cite{breiman2001random} or by the subsequent permutation of the feature values. While MDI feature importance depends on the training set and is consequently dependent on overfitting, permutation importance can be calculated subsequently based on the test set. The dataset parameters of our experimental setting are shown in table \ref{tab:basic_statistis} of the appendix. 

To evaluate the performance of the classifier and calculate the permutation importance, we use the area under the curve (AUC), as the distribution of the target variables is very skewed. For dataset 1 and 2, the random forest classifier reaches an AUC of 0.6163 and 0.6157, respectively. 
% TODO rewrite this + give explanation 
%Especially for dataset 2 the most important instances are those with a lower frequency in $S_{region}$. If we restrict the selection of the patterns to those with a higher frequency in $S_{region}$, the AUC decreases to 0.614 and 0.546.\\
In order to ensure that the classification performance cannot be obtained by a random selection, we rerun the experiment with a random SNP selection instead of the selected SNPs. In this setting, the random forest classifier only reaches an AUC of 0.534 and 0.525 for dataset 1 and 2, respectively. 
The most influential patterns are listed in table \ref{tab:most_important_pattern}. While the pattern frequency difference of dataset 1 correlates with the feature importance, this holds less on dataset 2, since also the pattern frequency difference is decreased.
%Interpretation?

\section{Conclusion}
\label{conclusion}
In this work, we have shown the feasibility of an unsupervised detection of selected SNPs under consideration of multivariate dependencies for Pool-seq data by the use of CNNs. %unsupervised multivariate
The suggested method is realized with an alignment discriminating CNN and object recognition techniques of explainable artificial intelligence.
Compared to univariate statistical applications, our pipeline is able to consider whole regions of an alignment affected by environmental adaptation. Our overall unsupervised method allows us to find selected SNPs without the need for a given ground truth, which increases the flexibility of our approach.
%During the direct comparison with SNPs found by univariate statistical methods we are able to find a  superset that prove the validity of both our two main assumptions. 
The direct comparison with SNPs found by univariate statistical methods demonstrates that our pipeline finds a superset of the compared results, which is in line with our main assumptions. 
We show that research on unsupervised selected SNP detection has the potential to find so far unknown regions of environmental adaptation. Therefore, our method offers the possibility to further supplement univariate statistical results, in cases where this is needed.
\section*{Acknowledgments}
This work was supported by the Research Center
For Emergent Algorithmic Intelligence at the University of Mainz funded by the Carl-Zeiss Foundation.
\bibliographystyle{unsrt}  
\bibliography{references}
\appendix
\section{Appendix}

\begin{table}[h]
 \caption{Absolute values of our selection method for different choices of $\lambda_{3}$. $S_{gene}$ and $R_{gene}$ are the corresponding sets of genes with selected SNPs. The number of genes of the univariate SNPs $|R_{gene}|$ are 1893 and 3803 for dataset 1 and 2, respectively. The first line shows the result of $|S|=|R|$.}
  \centering
  \begin{tabular}{lllll|lllll}
    \toprule
    \multicolumn{5}{c|}{Dataset 1}& \multicolumn{5}{c}{Dataset 2} \\
    $\lambda_{3}$&$|S|$&$|S-R|$&$|S_{gene}|$&$|S_{gene}-R_{gene}|$&$\lambda_{3}$&$|S|$&$|S-R|$&$|S_{gene}|$&$|S_{gene}-R_{gene}|$\\
    \midrule
    0.011&10539&10397&4891&4273&0.17&18256&14817&3208&1377\\ 
    0.075&74436&73301&11773&10358&0.075&8216&6607&2161&925\\ 
    0.1&99248&97759&12202&10744&0.1&10955&8860&2510&1083\\ 
    0.125&124060&122203&12444&10957&0.125&13693&11089&2804&1201\\ 
    0.15&148872&146663&12584&11080&0.15&16432&13325&3082&1321\\ 
    0.175&173684&171137&12672&11156&0.175&19171&15595&3270&1401\\ 
    0.2&198496&195645&12718&11197&0.2&21910&17866&3439&1471\\
    0.3&297744&293452&12852&11310&0.3&32865&26958&3952&1695\\ 
    0.4&396993&391441&12897&11344&0.4&43820&36143&4290&1833\\ 
    0.5&496241&489528&12914&11358&0.5&54775&45307&4493&1889\\ 
    0.6&595489&587892&12919&11362&0.6&65730&54390&4600&1909\\
    \bottomrule
  \end{tabular}
  \label{tab:absolute_values}
\end{table}

\begin{table}[h]
 \caption{Parameters of the empirical pattern detection for dataset 1 and 2.}
  \centering
  \begin{tabular}{lc|c}
    \toprule
    &Dataset 1&Dataset 2 \\
    \midrule
    Number of features&1000&141\\
    Training set size&29329&47652\\
    Test set size&19553&31769\\
    Number of positive instances&10539&18256\\
    Number of negative instances&38343&61165\\
    \bottomrule
  \end{tabular}
  \label{tab:basic_statistis}
\end{table}

\begin{table}[h]
 \caption{Top 30 of the most influencing patterns for each dataset according to the permutation importance of random forest classifier. Freq. 1 and freq. 2 represent the frequency of each pattern in $S_{region}$ and $P_{region}$, respectively. }
  \centering
  \begin{tabular}{llll|llll}
    \toprule
    \multicolumn{4}{c|}{Dataset 1}& \multicolumn{4}{c}{Dataset 2} \\
    Pattern&Freq. 1&Freq. 2& Importance&Pattern&Freq. 1&Freq. 2& Importance\\ 
    \midrule
    AAAAAAA&.192&.149&.000541&AAATAT&.152&.151&.002233\\
    AAAAAGT&.203&.169&.000854&AACTT&.257&.253&.002292\\
    AAAAATA&.334&.286&.000857&AAGAT&.215&.212&.002255\\
    AAAACG&.216&.175&.000793&AGTTC&.138&.137&.002177\\
    AAATAG&.288&.249&.000554&AOOA&.215&.213&.002637\\
    AACGC&.2633&.231&.000824&ATTAAT&.152&.147&.002777\\
    AATTAC&.272&.241&.000427&CAAAAT&.129&.128&.002257\\
    ACGCA&.261&.226&.000551&CAAAT&.294&.289&.00235\\
    ACGTA&.224&.188&.000544&CAGG&.177&.174&.00452\\
    ACTAG&.305&.284&.000418&CCAG&.204&.203&.002802\\
    ATACAA&.278&.235&.002146&CCTG&.173&.17&.003107\\
    ATACG&.184&.145&.000815&CGAT&.275&.273&.004301\\
    ATTCG&.162&.136&.000486&CGTC&.163&.154&.003747\\
    CAATTA&.267&.224&.001140&CGTT&.212&.209&.003528\\
    CCAAAG&.238&.216&.000432&CTGTA&.103&.096&.00261\\
    CCATTA&.178&.159&.000491&CTTAT&.174&.17&.002242\\
    CGAAT&.166&.139&.0009&GAGAT&.125&.122&.002853\\
    CGCAA&.186&.15&.000956&GCATT&.175&.166&.002888\\
    CGCAC&.16&.139&.000399&GTGAT&.142&.135&.002466\\
    CGCAT&.185&.162&.000471&OACT&.145&.144&.003197\\
    CGTAA&.226&185&.001112&OTCO&.131&.13&.003\\
    CTACG&.152&.126&.000411&TACAA&.198&.195&.002346\\
    GAAAATA&.209&.175&.000399&TATTA&.251&.249&.002888\\
    GACATA&.16&.14&.000505&TCACT&.134&.13&.003406\\
    TAAAATAA&.134&.109&.000431&TCAGC&.112&.106&.003061\\
    TAAAGTA&.108&.089&.000776&TGATC&.136&.13&.003176\\
    TAACAA&.302&.263&.000725&TTTAAT&.165&.161&.002222\\
    TACGT&.218&.186&.000464&TTTTC&.304&.3&.002184\\
    TCAAAAT&.182&.156&.000431&TTTTCA&.14&.132&.002706\\
    TTACG&.22&.186&.000571&TTTTGA&.137&.132&.002287\\
    \bottomrule
  \end{tabular}
  \label{tab:most_important_pattern}
\end{table}
\end{document}